\documentstyle[12pt,aaspp4,epsf,flushrt]{article}

\newcommand{\sig}{\:\lower0.6ex\hbox{$\stackrel{\textstyle >}{\sim}$}\:}
\newcommand{\sil}{\:\lower0.6ex\hbox{$\stackrel{\textstyle <}{\sim}$}\:}

\lefthead{Klessen, R.~S. and Burkert, A.}
\righthead{Fragmentation of Molecular Clouds}

\begin{document}

\title{On the Formation of Stellar Clusters:  Gaussian Cloud Conditions II}
    \author{Ralf S. Klessen$^{1,2}$ and Andreas Burkert$^{2}$\\ 
{$^{1}$\small Sterrewacht Leiden, Postbus 9513, 2300-RA Leiden,
    The Netherlands\\}
{$^{2}$\small   Max-Planck-Institut f\"{u}r Astronomie, K\"onigstuhl 17, 69117
    Heidelberg, Germany}}


\begin{abstract}
Using hydrodynamic simulations we investigate the time evolution and
fragmentation of regions within molecular clouds which have lost their
turbulent support leading to gravitational contraction. The initial
density distributions are described by random Gaussian fluctuations
with varying slopes $\nu$ of the power spectrum $P(k) \propto
k^{-\nu}$, covering the range from flat ($\nu=0$) to very steep
spectra ($\nu=3$). We consider molecular cloud volumes containing
different masses relative to the average Jeans mass $M_{\rm J}$, from
$1\,M_{\rm J}$ to $222\,M_{\rm J}$. This parameter study extends the
detailed analysis of systems with initially $P(k) \propto k^{-2}$ and
mass $222\,M_{\rm J}$ presented by Klessen \& Burkert (2000).

The dynamical evolution of the simulated molecular cloud regions is
insensitive to the slope of the initial density fluctuation
spectrum. The system evolves into a complex network of intersecting
filaments and collapsing clumps leading to the formation of a compact
cluster of accreting and interacting embedded protostellar cores. The
cluster builds up as bound entity, but dissolves later due to
collisional effects.  In all simulations, the mass spectrum of
collapsed cores is very broad, has approximately log-normal shape and
peaks roughly at the average Jeans mass. This supports the hypothesis
that the average Jeans mass is the main parameter determining the peak
in the stellar spectrum, and suggests that the interplay between
self-gravity on the one side and thermal and turbulent pressure on the
other side is the dominant process that regulates the formation of
stellar clusters.
\end{abstract}


\renewcommand{\thefootnote}{\fnsymbol{footnote}}

\section{Introduction}
\label{sec:introduction}

 Understanding the processes that lead to the formation of stars is
one of the fundamental challenges in astronomy. All presently known
star formation takes place in turbulent interstellar clouds of cold
molecular hydrogen. The formation of molecular clouds out of the
diffuse and warm atomic interstellar medium is not fully understood,
and probably involves large-scale compressible turbulence, chemical
phase transitions, and a combination of different instability
mechanisms including gravitational, thermal, Kelvin-Helmholtz and
Rayleigh-Taylor instabilities (e.g.\ Elmegreen 1993,
V{\'a}zquez-Semadeni et al.\ 2000 for an overview, or Burkert \& Lin
2000 for an analysis of the thermal instability).  Within the highly
structured and dynamically active molecular cloud environment, stars
form in groups and build up from gas that is subject to localized
gravitational instability. The density in the central part of a
collapsing gas clump increases enormously and a protostellar object
forms. The nascending protostar grows in mass via accretion from the
infalling envelope until the available gas reservoir is exhausted or
stellar feedback effects become important and remove the parental
concon --- a new star is born (for a comprehensive overview consult
Mannings, Boss \& Russell 2000, and references therein). This complex
evolutionary sequence involves a wide variety of different physical
phenomena, and it is not at all well understood which processes
dominate and determine the spectrum of stellar masses and the
kinematic and spatial properties of young star clusters.

The current investigation focuses on the initial phases of the star
formation process. We model the fragmentation and dynamical evolution
of large subvolumes within molecular clouds towards the formation of
clusters of protostellar cores using smoothed particle hydrodynamics
(SPH -- see the reviews of Benz 1990 and Monaghan 1992) implemented
with the special-purpose hardware device {\sc Grape} (Sugimoto et al.\
1990, Ebisuzaki et al.\ 1993). This extends previous studies of the
collapse of isolated gas clumps (e.g. Burkert \& Bodenheimer 
1996; Burkert, Bate \& Bodenheimer 1997; Boss 1997, Truelove et al.\
 1998, Bate 1998) to the molecular cloud regime and allows us to
consistently and statistically study the formation process of stellar
clusters within globally stable molecular clouds.  Stellar
cluster formation through gravitational collapse and fragmentation of
isolated gas spheres has been modeled with low resolution by Larson
1978, and Keto, Lattanzio \& Monaghan 1991. Star formation at the interface of
colliding gas spheres has been studied by Whitworth et al (1995),
Turner et al. (1995) and Bhattal et al. (1998), and the effect of gas
accretion onto embedded young stellar clusters was discussed by
Bonnell et al.\ (1997).  The simulations presented here combine all
these different aspects of star formation by self-consistently
following the isothermal evolution of turbulent molecular clouds
towards the formation of stellar clusters including its competitive
accretion phase.

As we want to understand which physical effects  determine the
properties of stellar clusters and how these depend on different
initial cloud conditions, we adopt the following approach: It is known
from numerical studies that interstellar turbulence decays very
rapidly, on time scales of the order of the free-fall time scale of
molecular clouds (Gammie \& Ostriker 1996, Mac~Low et al.\ 1998,
Stone, Ostriker \& Gammie 1998, Mac~Low 1999, Padoan \& Nordlund
1999). This is also the typical time scale inferred for the formation
of star clusters (e.g.\ Carpenter et al.\ 1997, Hillenbrand \&
Hartmann 1998, also Elmegreen 2000). Clustered star formation
therefore occurs in molecular cloud regions which have (locally) lost
their turbulent pressure support or, almost equivalently, which are only
weakly supported by turbulence on large scales (Klessen, Heitsch \&
Mac~Low 2000).  The turbulent
velocity field in the subvolume under consideration
is assumed to leave a Gaussian fluctuation spectrum as
imprint into the density distribution of the cloud.  The dynamical evolution
of the cloud in the isothermal phase is thus determined by two
parameters, the power spectrum $P(k)$ of the fluctuations and the gas
temperature.  We consider power-law spectra $P(k) \propto k^{-\nu}$
with slopes ranging from $\nu=0$ to $\nu=3$ and gas temperatures such
that the considered volume contains between 1 and 222 thermal Jeans
masses.  In the first part of this investigation (Klessen \& Burkert
2000, in the following denoted Paper I; see also Klessen, Burkert, \&
Bate 1998) we presented the results of a
detailed case study and focused on an in-depth analysis of the
formation of stellar clusters in subregions containing 222 Jeans
masses within clumpy molecular clouds with power spectrum $P(k)
\propto k^{-2}$. Using some of the statistical tools introduced in
Paper I, we discuss here selected aspects of the complete survey of
the parameter space which we performed.

The results of Paper I showed that even simple, isothermal models of
self-gravitating clumpy clouds are able to explain many of the
observed features of star-forming regions. This is rather surprising,
given the fact that magnetic fields and energetic heating processes of
newly formed stars could in principle be important, both of which have
been neglected. The early stages of star formation are dominated by
the interplay between turbulence and gas pressure on one side and the
gravitational forces on the other.  This creates an intricate network
of filaments, sheets and dense clumps. Some clumps will become
gravitationally unstable and undergo rapid collapse. While contracting
individually to form protostellar cores in their interior, gas clumps
stream towards a common center of attraction. Throughout this paper,
the terms gas clump and density fluctuation are used synonymously,
cores are defined as the high-density central regions of collapsing
clumps where individual protostars build up.  As described in Section
2, the numerically unresolved cores are replaced by sink
particles. Altogether, the dynamical evolution of molecular clouds
involves processes acting simultaneously on different length scales
and time scales. A quasi-equilibrium clump mass spectrum emerges and
follows a power-law $dN/dM \propto M^{-1.5}$. Individual clumps are
elongated, centrally condensed objects with 2:1 to 4:1 axis ratios and
with $r^{-2}$ density fall-off.  In contrast to the clumps, the core
mass spectrum is best described by a log-normal distribution with a
peak and a width that is in excellent agreement with observations of
multiple stellar systems.

The presence of unpredictable dynamical processes in the overall gas flow
and the evolution of the nascending protostellar cluster very
efficiently erases the memory of the initial configuration.  For this
reason, it is impossible to predict the detailed evolution of
individual objects from the initial state of the system. Only the
properties of an {\em ensemble} of protostellar cores, for example
their kinematics and mass distribution, can be determined in a
probabilistic sense. A comprehensive theory of star formation needs to
be a statistical theory.

In the present part II, we investigate how the formation of
protostellar clusters depends on variations in the environmental and
initial conditions.  The structure of the paper is as follows. First
(in \S\ref{sec:models}) we briefly recapitulate the properties of our
numerical models as introduced in detail in Paper I, then we analyze
(\S\ref{sec:dep-power-spectrum}) the dependence of the dynamical
evolution of the system on the slope $\nu$ of the initial fluctuation
spectrum. We discuss the stability of the resulting core clusters in
\S\ref{sec:dynamical-stability}, and vary the temperature in
\S\ref{sec:dep-temperature}. Finally, we summarize our results in
\S\ref{sec:parameter-study-summary}.

\section{Models}
\label{sec:models}

\subsection{SPH in Combination with GRAPE}
\label{subsec:SPH}
SPH ({\em smoothed particle hydrodynamics}) is a Lagrangian method to
solve the equations of hydrodynamics. The fluid is represented by an
ensemble of particles and thermodynamical observables are obtained by
averaging over an appropriate subset of the SPH particles (Benz 1990,
Monaghan 1992). The method is is very flexible and able to resolve
very high density contrasts by increasing the particle concentration
where needed.  We use SPH in combination with the special-purpose
hardware device GRAPE (Sugimoto et al.~1990, Ebisuzaki et al.~1993),
which allows calculations at supercomputer level on a normal
workstation. In addition, we implemented periodic boundary conditions
(Klessen 1997) to describe molecular clouds that are stable against
gravitational collapse on global scales and to concentrate on the
dynamical evolution of subvolumes within those clouds.

In any self-gravitating fluid, regions with masses exceeding the Jeans
limit become unstable and collapse. In our implementation of SPH, once
a highly-condensed object forms in the center of a collapsing gas
clump and has passed beyond a certain density threshold, the dense
core is substituted by a sink particle (Bate, Bonnell \& Price
1995). The density limit is four to five orders of magnitude above the
initial average density, depending on the particle number in the
simulation, and the accretion radius of the sink particle is slightly
larger than the corresponding Jeans length. The particle inherits the
combined masses, linear and relative angular momenta of the SPH
particles it replaces, and has the ability to accrete further SPH
particles from its infalling gaseous surrounding. Again, the mass,
linear and angular momenta are conserved. By adequately replacing
high-density cores by sinks and keeping track of their further
evolution in a consistent way we are able to follow the dynamical
evolution of the system over many free-fall times.

More details about the implementation of SPH which we use and its
advantages and limitations of the method are discussed in \S2 of
Paper I.

\subsection{Initial Conditions}
\label{sec:scaling}
We investigate the first phases of star formation, the dynamical
evolution and fragmentation of structure within molecular clouds
towards the formation of clusters of embedded accreting protostellar
cores. In this regime the gas is approximately isothermal with a
typical temperature of $10\,$K (e.g.\ Tohline 1982). Stellar clusters
are known to form quickly on the order of the local collapse time
scale (e.g.\ Hillenbrand \& Hartmann 1998, Elmegreen 2000). This
requires the coherent loss of turbulence within a sufficiently large
molecular cloud region and we start our simulations at the onset of
gravitational contraction. The initial density fluctuation spectra are
taken to be Gaussian with power distribution $P(k) \propto k^{-\nu}$,
with the exponents in the range $\nu=0$ to $\nu=3$. The velocity is
computed self-consistently from the density by solving Poissons
equation, i.e.\ the gas velocity at any location is due to
gravitational contraction only, additional contribution from
interstellar turbulence are assumed to have decayed away.

The slope $\nu=0$ of the density spectrum implies equal power on all
scales. As a result, the structure of the system appears relatively
homogeneous without large-scale features. Such density structure is
the imprint of turbulence that contained most energy on small
scales. As the exponent $\nu$ increases, the fluctuation spectra become
steeper and the spatial structure of the simulated molecular clouds
volume become more and more dominated by large-scale modes. This
effect is nicely illustrated in figure \ref{fig:cube-3D-T01Nx} at
$t=0$. Large-scale density fluctuations are assumed to stem from
turbulent velocity fields that have been dominated by large-scale
modes.

Assuming that stellar clusters form through the gravitational collapse
of clumpy cloud regions that have lost their turbulent support, we
study the detailed behavior of this process and the properties of the
newly formed cluster. The current study does not include the possible
effect of magnetic fields on this process.  However, the overall
importance of magnetic fields and MHD waves on the dynamical structure
of molecular clouds may not be large. The energy associated with the
observed fields is of the order of the (turbulent) kinetic energy
content of molecular cloud clumps, and smaller than the gravitational
one (Crutcher 1999). Hence, the observed fields are too weak to
prevent local collapse. In addition, the presence of magnetic fields
cannot halt the decay of turbulence (Mac~Low et al.~1998, Stone et
al.\ 1998, Padoan \& Nordlund 1999), and may not be able to
significantly alter the efficiency of local collapse in the case of
driven turbulence (Heitsch, Mac Low, \& Klessen 2001).  In the case of
stable cloud regions external driving mechanisms are necessary,
otherwise kinetic energy dissipates quickly in regions without driving
sources, resulting in the rapid formation of stellar clusters. This is
what we consider here. Star formation on a crossing time scale is
implied by observational evidence also on scales of molecular clouds
as a whole (Ballesteros-Paredes, Hartmann, \& V\'azquez-Semadeni 1999,
Elmegreen 2000), again supporting the present model where magnetic
fields does not play a dominant role.
 
We adopt random Gaussian fluctuations for the density as starting
condition for the SPH simulations, because their properties resemble
the end stages of decaying turbulence and are suited to mimic observed
features of molecular clouds once advanced into the non-linear regime
(see e.g.~Stutzki \& G{\"u}sten 1990, who deploy a Gaussian
decomposition technique to describe the clumpy structure of molecular
clouds). It is important for the current parameter study, that these
distributions have well defined statistical properties.  Gaussian
random fields are completely characterized by the normalization and
the power spectrum $P(\vec{k})$, i.e.\ all higher moments can be
expressed in terms of the first two moments.  In case of isotropy, the
power distribution is independent of direction and depends only on the
wave number $k=|\vec{k}|$.  The function $P(k)$ identifies the
contribution of waves with wave number $k$ to the statistical
fluctuation spectrum.  The phases of Gaussian random fields are
arbitrarily chosen from a {\em uniform} distribution in the interval
$[0,2\pi[$, and the amplitudes for each mode $k$ are randomly drawn
from a {\em Gaussian} distribution with width $P(k)$ centered on
zero. Since waves are generated from random processes, only the
properties of an {\em ensemble} of fluctuation fields are determined
in a statistical sense. Individual realizations (from different sets
of random numbers) may deviate considerably from this mean value,
especially at small wave numbers $k$.  We generate the initial density
fluctuation fields by applying the Zel'dovich (1970) approach. The
method and its applicability to gaseous systems is discussed in detail
in \S3 and Appendix B of Paper I.

\subsection{Scaling Properties}
\label{subsec:scaling-properties}
For isothermal gas, the energy density is a function of temperature
only and the equation of state reduces to $p = c^2_{\rm s}\;\!  \rho$,
with $c_{\rm s}$ being the thermal sound speed.  The self-gravitating,
isothermal model studies the interplay between gravity and gas
pressure. Given the initial density and velocity distribution, i.e.\ a
slope $\nu$ of the density fluctuation spectrum, the dynamical
evolution of the system depends only on one free scaling parameter,
the ratio $\alpha$ between the total internal energy $\epsilon_{\rm
int}$ and the total potential energy $\epsilon_{\rm pot}$ in the
system, which is equivalent to a dimensionless {\em temperature},
\begin{equation}
\label{eqn:def-temperature}
\alpha \equiv \epsilon_{\rm int}/|\epsilon_{\rm pot}|\:.
\end{equation}
Generally, all physical parameters and constants in our models are
normalized and set to unity. The same applies to mass and length
scales. The total mass in the system is $M=1$ and the simulated volume
is the cube $[-1,+1]^3$, i.e.\ the mean density is $\langle \rho
\rangle = 1/8$.  In these units the thermal Jeans mass for spherical
perturbations is
\begin{equation}
\label{eqn:def-jeans-mass}
M_{\rm J} = 1.6\cdot\rho^{-1/2}\;\alpha^{3/2}\;.
\end{equation}
Using the average value $\langle \rho \rangle = 1/8$ for the density,
the global free-fall time is $\tau_{\rm ff}= 1.5t$ with $t$ being the
dimensionless time unit. The parameter $\alpha$ regulates the number
of thermal Jeans masses contained within the simulated volume, $N_{\rm
J} = 1/M_{\rm J}$.  We vary the dimensionless temperature from $\alpha
= 0.01$ to $\alpha = 0.5$. In Paper I we considered $\alpha = 0.01$,
which corresponds to $N_{\rm J}=222$ and leads to the formation of
clusters of $\sim 60$ protostars. A value $\alpha = 0.5$ implies
$N_{\rm J} = 0.6$. Collapse does not occur. When increasing the
scaling parameter $\alpha$, the total number of Jeans masses contained
in the computed cube decreases. In the isothermal models this is
equivalent to `zooming' in onto smaller spatial volumes.  By this,
individual protostellar clumps are better resolved with the caveat of
neglecting the influence of modes with wave lengths larger than the
box size.

To illustrate the corresponding physical scaling, we consider sites of
low-mass star formation, like Taurus, with $n({\rm H}_2) \approx
10^2\,{\rm cm}^{-3}$ and $T\approx10\,$K.  Assuming a mean molecular
weight of $\mu = 2.36$, for $\alpha =0.01$ the simulated cube contains
the mass $M=6\;\!300\,$M$_{\odot}$ and has a size $L = 10.2\,$pc.  For
$\alpha = 0.5$ the volume is much smaller with $M=17.4$M$_{\odot}$ and
$L = 1.44\,$pc. The time unit is equivalent to
$t=2.2\times10^6\,$years and the average Jeans mass transforms to $
M_{\rm J} = 28\,{\rm M}_{\odot}$. When applied to dense, massively
star-forming clouds with typically $n({\rm H}_2) \approx 10^5\,{\rm
cm}^{-3}$ and $T\approx10\,$K, similar to the BN region in Orion, the
simulated cube translates into the mass $M = 200\,$M$_{\odot}$ and the
size $L = 0.32\,$pc for $\alpha = 0.01$ and $M=0.6\,$M$_{\odot}$ and
the size $L = 0.05\,$pc for $\alpha = 0.5$. The time unit now converts
to $t=7.0\times10^4\,$years and the mean Jeans mass for the
homogeneous distribution is $M_{\rm J} = 0.9\,{\rm M}_{\odot}$.

\section{Dependence on the Power Spectrum}
\label{sec:dep-power-spectrum}
In this section we analyze the dependence of the fragmentation of a
region inside a molecular cloud and of the properties of the
protostellar cluster that forms during the dynamical evolution on the
choice of the initial fluctuation spectrum $P(k) \propto 1/k^\nu$. Again
for models with $\alpha=0.01$ which contain 222 Jeans masses, we compare
the evolution of the system for different values of $\nu$, ranging from
$\nu=0$ which means that fluctuations of all wave lengths $k$ have
statistically the same amplitude, to the very steep power spectrum
$\nu=3$ which implies strong dominance of the large-scale modes. For
$\nu=0$, the initial density distribution looks quite homogeneous,
whereas it is strongly biased towards having one dominant density peak
in the case of $\nu=3$. Additionally, for $\nu=1$ we study the case of a
truncated spectrum where fluctuations on scales $k<4$ have been
removed.  For a comparison of the initial density fields with
different power spectra see Fig.~\ref{fig:cube-3D-T01Nx}.  The
parameters of all models discussed in this section are listed in
Tab.~\ref{tab:models-N=0,1,2,3}.

\subsection{Dependence on the Slope of the Power Spectrum}
\label{subsec:dependence-on-slope}
\begin{figure}[p]
\unitlength1.0cm
\hspace{-0.3cm}
\begin{picture}(16,19.7)
\put( 0.9, 0.0){\epsfbox{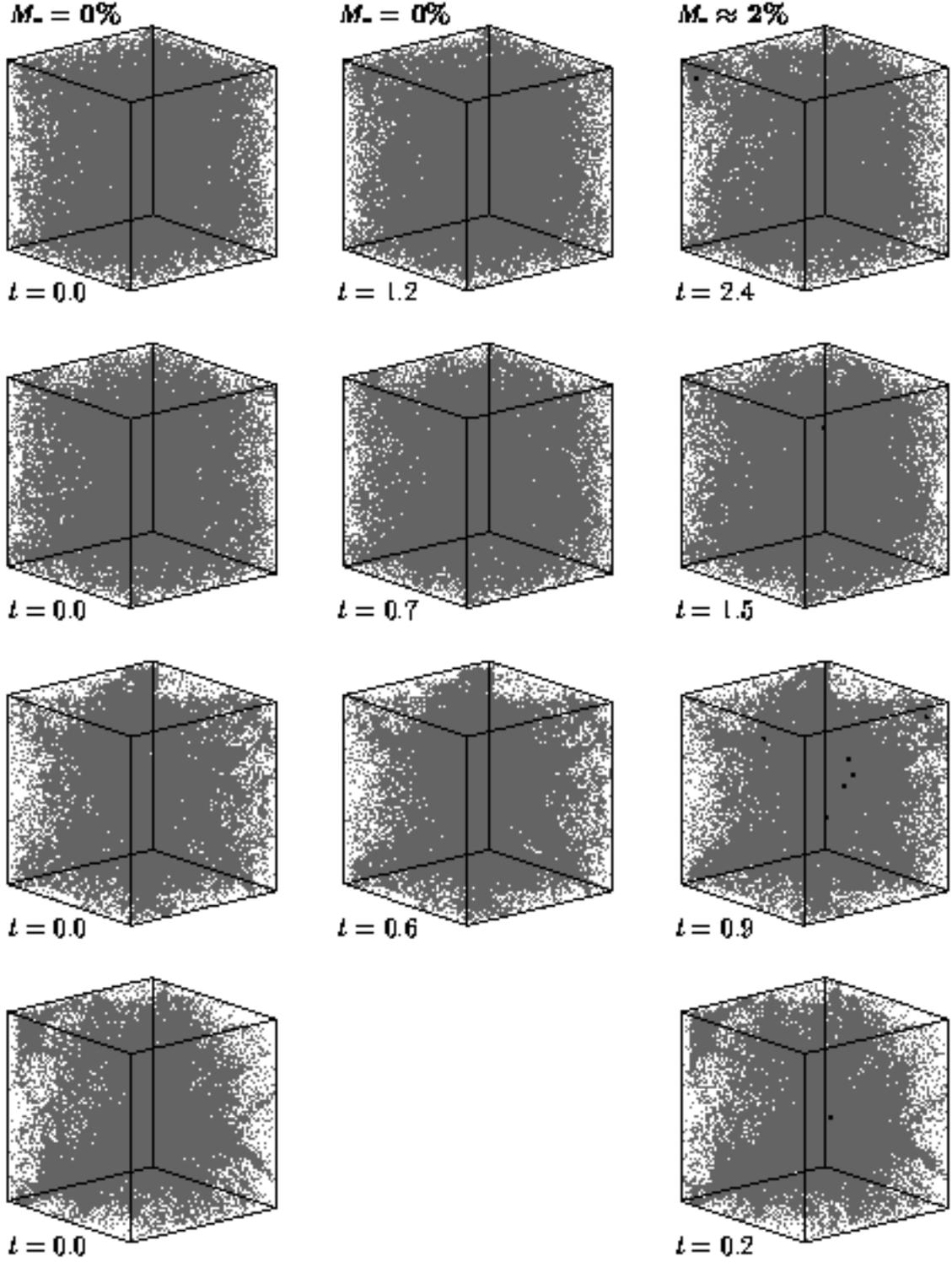}}
\end{picture}
\caption{\label{fig:cube-3D-T01Nx} \footnotesize Comparison of the time evolution of
four models with different initial power spectra $P(k) \propto 1/k^\nu$,
initially (first column), after the first protostellar cores have
formed (second column) and when $M_*\approx 2\,$\% of the gas is
converted into condensed cores.  The first row plots model ${\cal A}0a$
with $\nu=0$, the second row model ${\cal A}1a$ with $\nu=1$, the third row
model ${\cal A}2a$ with $\nu=2$, the fourth row model ${\cal A}3a$ with
$\nu=3$. }
\end{figure}

To examine how the variation of the slope $\nu$ of the initial
fluctuation spectrum influences the dynamical evolution of the gas
system, we generate four models with varying $\nu$ but otherwise
almost identical properties. Models ${\cal A}0a$, ${\cal A}1a$, ${\cal
A}2a$ and ${\cal A}3a$ have $\nu=0$, $\nu=1$, $\nu=2$ and $\nu=3$,
respectively. All consist of $2\times10^5$ SPH particles and are
generated via the Zel'dovich method from the same initial random
particle distribution, applying the same shift interval $\delta t =
1.5$.  To exclude the variance effects inevitable when comparing
different realizations of a fluctuation field with given statistical
properties, models ${\cal A}0a$, ${\cal A}1a$ and ${\cal A}3a$ are
computed from the same set of random numbers used to generate the
fluctuation spectrum. This means, that individual modes $k$ in each of
the three fields have identical {\em phases}. However, they differ in
{\em amplitude}, since these are drawn from Gaussians with different
width $P(k) \propto 1/k^\nu$, where $\nu=0$, 1 or 3, respectively. For
this reason, at comparable stages of their dynamical evolution, the
three models look remarkably similar. For model ${\cal A}2a$ a
different set of random numbers is used, and statistical variance
leads to a different appearance. This can be seen in
Fig.~\ref{fig:cube-3D-T01Nx}, which plots the 3-dimensional particle
distribution for each model at different stages of the dynamical
evolution. The first row denotes model ${\cal A}0a$, the second row
model ${\cal A}1a$, the third one model ${\cal A}2a$ and finally the
fourth row model ${\cal A}3a$. Each column in the figure shows the
distributions at comparable evolutionary phases, characterized by the
mass fraction $M_*$ accreted onto protostellar cores, as indicated at
the top of each column. The first column shows the initial density
distribution, the second column describes the state of the system when
the maximum density contrast has reached half the value necessary for
a collapsing object to be identified as protostellar core (see
Sec.~\ref{subsec:SPH}), the third column shows the system when 2\% of
the gas mass is contained in condensed cores, and so forth.

The overall dynamical evolution of all models is very similar: The
cloud develops a complex filamentary structure, and dense cores form
typically at the intersection of filaments. They accrete from the
surrounding gas while streaming towards the common center of
attraction.  It is common to all models to form {\em clusters of
protostellar cores which grow in mass via competitive accretion from
the common gas reservoir}, as discussed in detail in Paper I.

However, there are differences in the detailed behavior. It is visible
in the left column of Fig.~\ref{fig:cube-3D-T01Nx} that steeper
spectral slopes $\nu$ lead to initial density distributions that
exhibit more spatial structure and higher degrees of
inhomogeneity. The density fields of the models with $\nu \sig 2$ are
dominated by the largest-scale modes. On the other hand, for $\nu \sil
1$, fluctuations on smaller scales have sufficient amplitudes to
compensate the large-scale fluctuations when the whole spectrum is
added up, hence, model ${\cal A}0a$ (with $\nu = 0$) appears relative
homogeneous initially.  All initial fluctuation fields are generated
using the Zel'dovich method with constant shift interval $\delta t =
1.5$. As discussed in Appendix B of Paper I, this implies that the
initial density contrast increases with slope $\nu$, lying between
$\delta \rho/\rho = 3$ for ${\cal A}0$ and $\delta \rho/\rho = 50$ for
${\cal A}3$. Regions of higher density contrast begin to contract
faster and the system reaches a given evolutionary stage at earlier
times, since the free-fall time is a function of density, $\tau_{\rm
ff} \propto \rho^{-1/2}$.  Compared to model ${\cal A}3a$ with the
largest initial density contrast, model ${\cal A}0a$ lags behind by
$\Delta t \approx 2.5$, model ${\cal A}1a$ by $\Delta t \approx 1.5$
and model ${\cal A}2a$ by $\Delta t \approx 1.0$. Despite this initial
lag, the dynamical evolution runs synchronously, i.e.\ the time
intervals between different stages of star formation are comparable in
all models. The overall age spread between the appearence of the first
and the last identified cores is roughly 1.3 free-fall times
$\tau_{\rm ff}$.

The slope of the power spectrum and the initial density contrast also
influences the total number of protostellar cores that form during the
dynamical evolution. If superposed on large-scale density modes, the
probability of small-scale fluctuations to exceed the local Jeans
limit is increased by the contribution of the large-scale mode, since
$M_{\rm J}\propto \rho^{-1/2}$ (Eqn.\ \ref{eqn:def-jeans-mass}).  This
effect is not large for models ${\cal A}0a$ to ${\cal A}2a$, they all
form between 45 and 59 cores, but becomes important for model ${\cal
A}3a$. It forms altogether 130 sinks. 

\begin{figure}[p]
\unitlength1.0cm
\hspace{-0.3cm}
\begin{picture}(16,17.5)
\put(-2.0,-6.5){\epsfbox{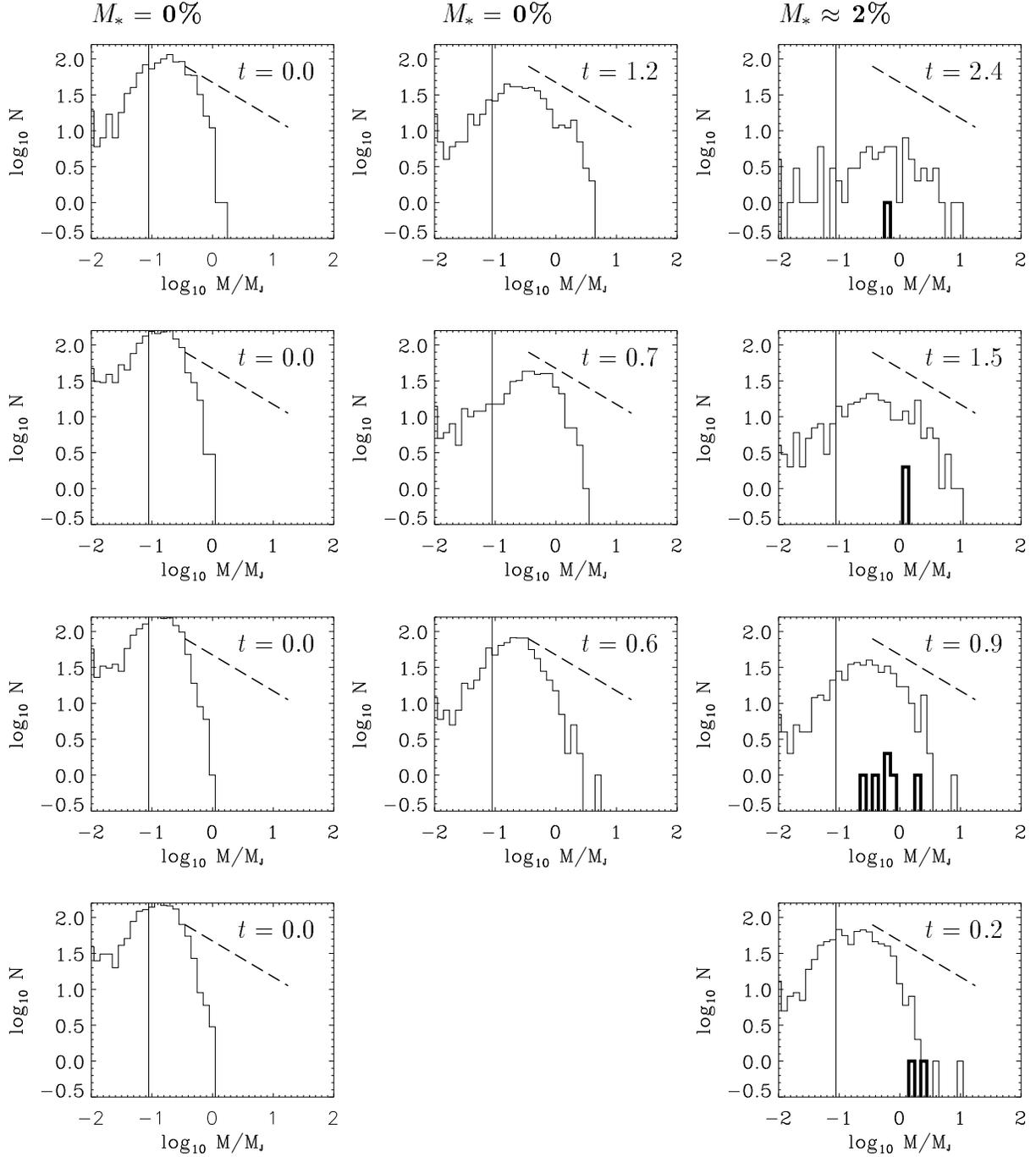}}
\end{picture}
\vspace{0.7cm}
\caption{\label{fig:mass-spectra-T01Nx} \footnotesize Mass spectrum of gaseous
clumps (thin lines) and protostellar cores (thick) lines in models
${\cal A}0a$ (first row), ${\cal A}1a$ (second row), ${\cal A}2b$
(third row), and ${\cal A}3a$ (fourth row). Each column describes
comparable evolutionary stages as defined by the fraction of mass
converted into condensed cores $M_*$ (analog to
Fig.~\ref{fig:cube-3D-T01Nx}). The corresponding time is indicated
separately in each plot. The thick dashed lines indicate the observed
clump mass spectrum $dN/dm \propto m^{-1.5}$, whereas the horizontal
lines give the resolution limit for the clump masses. }
\end{figure}

The mass distribution of identified gas clumps and protostellar cores
at comparably evolutionary stages are very similar. This can be seen
in Fig.~\ref{fig:mass-spectra-T01Nx}, which plots the number of gas
clumps (thin lines) and of protostellar cores (thick lines) as
function of their mass. Individual masses are scaled relative to the
Jeans mass of the homogeneous cube which is $M_{\rm J} =
1/222$. Analogous to Fig.~\ref{fig:cube-3D-T01Nx}, each row denotes an
individual model (from $\nu=0$ to $\nu=3$) and each column indicates
the same evolutionary phases as in the previous figure. The vertical
line denotes the SPH resolution limit, below which the local Jeans
mass is not resolved properly (see Bate \& Burkert 1997). The dashed
line indicates the observed slope of the clump-mass spectrum $dN/dm
\propto m^{-1.5}$.

Clearly, the initial conditions fail to reproduce the observed
clump-mass distribution. Our clump-finding algorithm detects the
initial Gaussian fluctuation spectrum and the clump spectra for the
different models appear identical since the algorithm is biased
towards small scale fluctuations: if a variety of small-scale peaks
are superposed on a large-scale mode, it breaks the latter one up into
the contributions of the smaller fluctuations. The large mode is
identified only if it is sufficiently `smooth', i.e.~if it cannot be
identified as the sum of an nsemble of smaller clumps. The algorithm
is similar to the one decribed by Williams, De\ Geus, \& Blitz (1994)
adopted to make use of the SPH kernel smoothing and thus differs from
the Gaussian decomposition scheme introduced by Sutzki \& G{\"u}sten
(1990). For a detailed description see Appendix A in Paper I.

As the system begins to evolve, gas pressure leads to the
disintegration of small clumps with masses below a Jeans mass, whereas
gravitational attraction causes more massive clumps to merge thus
creating larger ones with increasing mass. As a result, for models
${\cal A}0a$ and ${\cal A}1a$, the clump spectrum becomes flatter in
the interval $M_* \approx 1$\% to $M_*\approx30$\%. At later stages,
additional low-mass clumps are identified which may result from
irregularities or sub-fragmentation in the converging gas flows at the
intersection of two filaments along which gas streams towards a common
center of attraction (see Paper I for more details of this
processes). At very late stages, when almost all mass is contained in
protostellar cores, fitting a single power-law slope to the mass
spectrum becomes completely meaningless.  Since the models ${\cal
A}2a$ and ${\cal A}3a$ pass through all evolutionary stages at earlier
times, self-gravity and gas pressure have less time to shape the
clump-mass spectrum. Therefore, these systems contain more low-mass
clumps compared to the other two runs and their clump-mass spectrum is
on average steeper. Again at late stages the distribution can no
longer be fitted by a simple power law.

In summary, the clump-mass spectrum of self-gravitating isothermal gas
evolves in time and exhibits a well defined power-law behavior only
during the intermediate stages of its dynamical evolution. During this
period, the slopes vary $dN/dM$ gradually from values of about $-1$ to
$-2$. This is consistent with the variety of different slopes quoted
for observed clump-mass spectra and the uncertainty in their
determination (e.g.\ Stutzki \& G{\"u}sten 1990, Williams et al.\
1994, Kramer et al.\ 1998).

The mass distribution of protostellar cores is approximately
log-normal covering two orders of magnitude in mass. It peaks roughly
at the {\em mean Jeans mass} of the system. This is in agreement with
the results of Paper I and results from the fact that the formation of
new low-mass protostellar cores and the accretion onto already
existing ones are approximately in balance; they populate the low-mass
and high-mass side of the distribution more or less equally.  The peak
of the distribution moves away from the Jeans mass only at late stages
of the evolution, when the formation of new cores has stopped, but the
already existing ones are still able to grow in mass. This effect is
largest for model ${\cal A}0a$, since it forms the lowest total number
of condensed cores. Model ${\cal A}3a$ builds up three times more
cores. Therefore, each protostellar core is on average three times
lighter than in the other cases and the distribution peaks at smaller
masses. Only towards the end of the simulation, when almost all gas is
accreted onto the protostellar cores the peak reaches values of the
average Jeans mass.

\subsection{Dependence on the Maximum Wave Number}
\label{subsec:dependence-k}
The initial conditions for the SPH simulations are determined by the
slope of the initial fluctuation spectrum. Furthermore, they depend on
the number of modes that contribute to the overall fluctuation
field. The initial fluctuations discussed so far are all generated
from waves with $k=1$, where the wave length is equal to the total
size of the considered volume, down to $k=32$, where the wave fits 32
times into the considered cube. For steep power spectra, the overall
appearance of the fluctuation field is strongly dominated by the
presence of the largest-scale modes (see
Fig.~\ref{fig:cube-3D-T01Nx}). To investigate the importance of these
modes on the evolution, model ${\cal A}1b$ has a initial a fluctuation
spectrum truncated for modes $k<4$ with slope $\nu =1$. It is
generated with a Zel'dovich shift interval $\delta t= 1.0$.

\begin{figure}[t]
\unitlength1.0cm
\hspace{-0.3cm}
\begin{picture}(16,14.3)
\put( 0.0, 0.0){\epsfbox{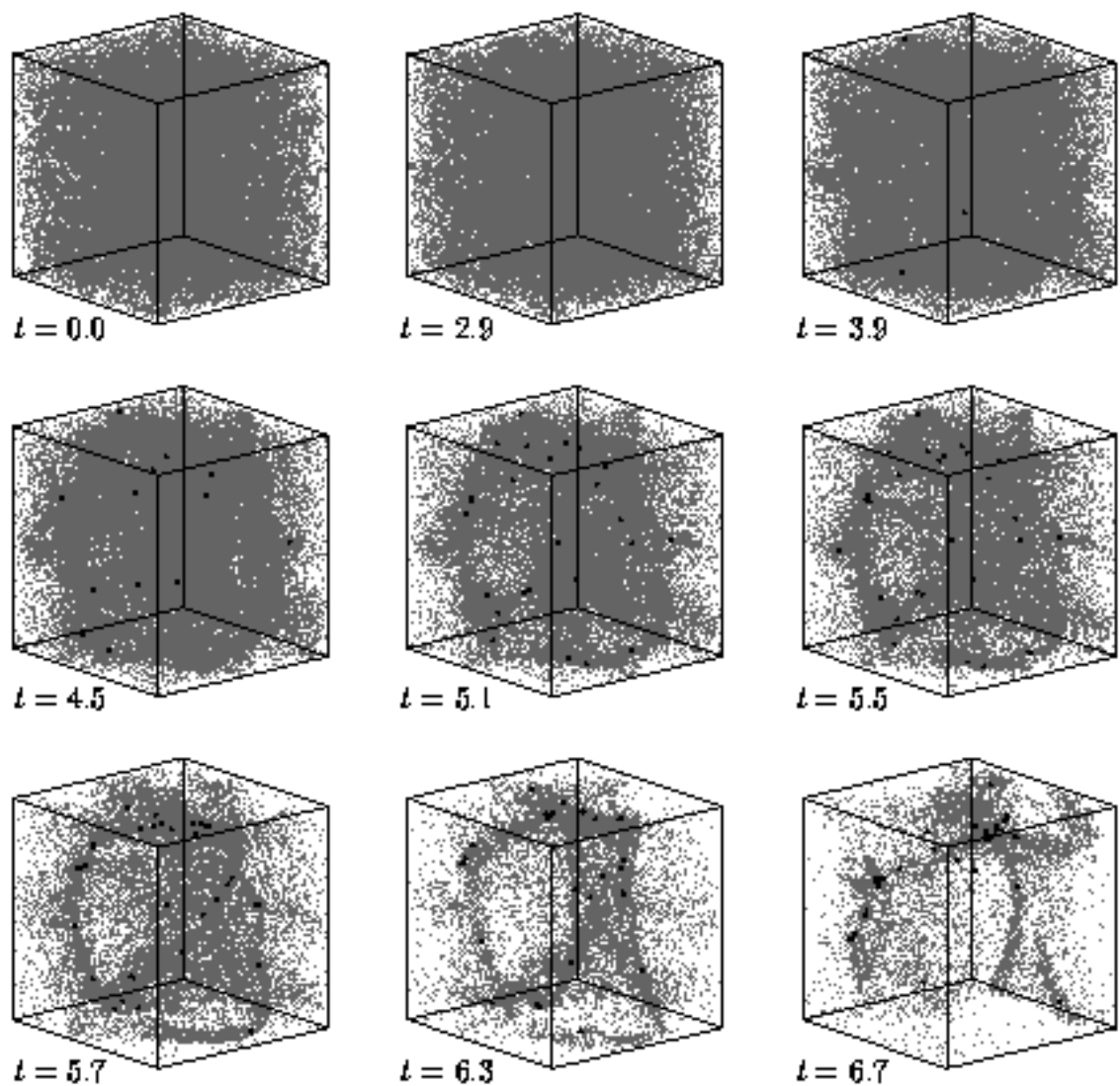}}
\end{picture}
\caption{\label{fig:cube-3D-T01N1-C} \footnotesize Time evolution of model ${\cal
A}1b$. It contains $5\times10^5$ SPH particles and the initial density
distribution is generated from a fluctuation spectrum with $\nu=1$
that is truncated for modes $k<4$. The snapshots correspond to the
following stages of the dynamical evolution: $t=0.0$ -- initial
particle distribution, $t=2.9$ -- the maximum density contrast has
reached half the value required to identify compact objects as
protostellar cores, $t=3.9$ -- the first protostellar cores have
formed and contain altogether $M_* = 2$\% of the total gas mass,
$t=4.5$ -- $M_* =10$\%, $t=5.1$ -- $M_* =30$\%, $t=5.5$ -- $M_*
=50$\%, $t=5.7$ -- $M_* =60$\%, $t=6.3$ -- $M_* =75$\%, and $t=6.7$ --
$M_* = 85$\%.  For legibility, only every tenth non-accreted gas
particle is displayed (small gray dots). Protostellar cores are
denoted by large dark dots.
}
\end{figure}

The time evolution of the system is displayed in
Fig.~\ref{fig:cube-3D-T01N1-C}, which shows the particle distribution
at different stages of the dynamical evolution characterized by the
total mass fraction $M_*$ accumulated in protostellar cores. Since the
Zel'dovich shift interval is relatively small and the initial
fluctuation field lacks the presence of large-scale modes, the
particle distribution appears very smooth and homogeneous. Its
evolution is most comparable to model ${\cal A}0a$, which also is very
homogeneous and smooth at the beginning of the evolution.  The initial
density contrast in model ${\cal A}1b$ is even smaller than in model
${\cal A}0a$, which implies that it needs longer to reach comparable
evolutionary stages, e.g.~it takes $t\approx 3.9$ to form the first
highly-collapsed protostellar object.  However, as in all other
models, a bound cluster of protostellar cores builds up. Again, the
mass spectrum is very broad, spanning two orders of magnitude, and
peaks slightly above the mean Jeans mass as derived from the mean
density; the mass spectrum of protostellar cores is indistinguishable
from the one of model ${\cal A}0a$.

 In summary, removing the large-scale modes is equivalent to flattening
the overall spectral slope. The details depend on the slope of the
initial spectrum and the wave number up to which fluctuations are
deactivated. Removing all modes with $k<4$ in a model with $\nu=1$
produces a system that in many aspects behaves like a model with
$\nu=0$.

\section{The Dynamical Stability of the Protostellar Cluster}
\label{sec:dynamical-stability}
\begin{figure}[t]
\unitlength1.0cm
\begin{picture}(16,11.7)
\put(-2.5,-9.0){\epsfbox{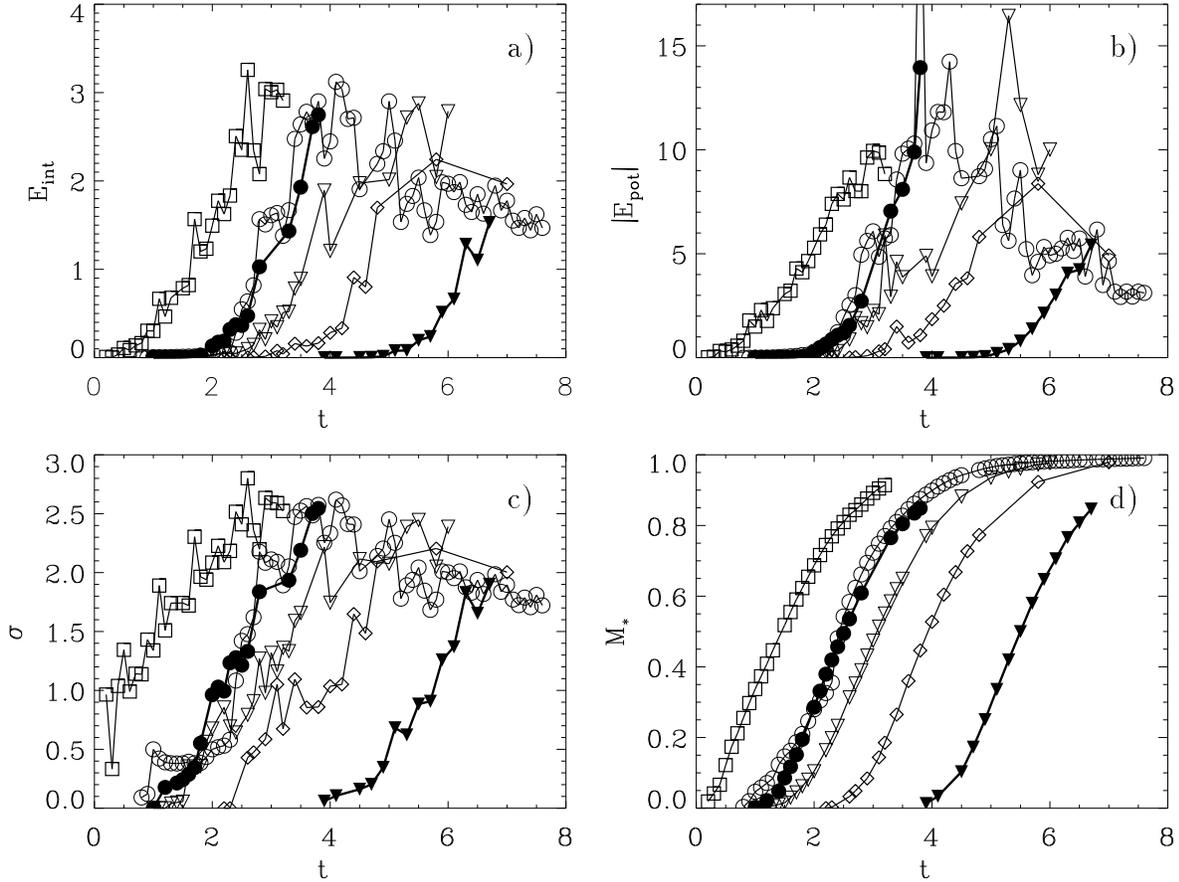}}
\end{picture}
\caption{\label{fig:energy-1} \footnotesize Time evolution of (a) the kinetic energy
in random motions $E_{\rm int}$, (b) the potential energy $E_{\rm
pot}$, (c) the velocity dispersion $\sigma$, and (d) the cumulative
mass $M_*$ in condensed cores for the clusters of protostellar cores
that build up. The open diamonds denote model ${\cal A}0a$ with
$\nu=0$. There are two models with $\nu=1$: ${\cal A}1a$ which is
denoted by the open triangles and the high-resolution model ${\cal
A}1b$ which is plotted with filled triangles (its initial fluctuation
spectrum is truncated and contains only modes with $k\ge4$). The
circles denote models with $\nu=2$: ${\cal A}2a$ (open circles) and
the high-resolution model ${\cal A}2b$ (closed circles). Finally,
model ${\cal A}3a$ with the steepest spectrum, $\nu=3$, is
characterized by open squares.  }
\end{figure}

\begin{figure}[t]
\unitlength1.0cm
\begin{picture}(16,17.5)
\put(-2.5,-6.0){\epsfbox{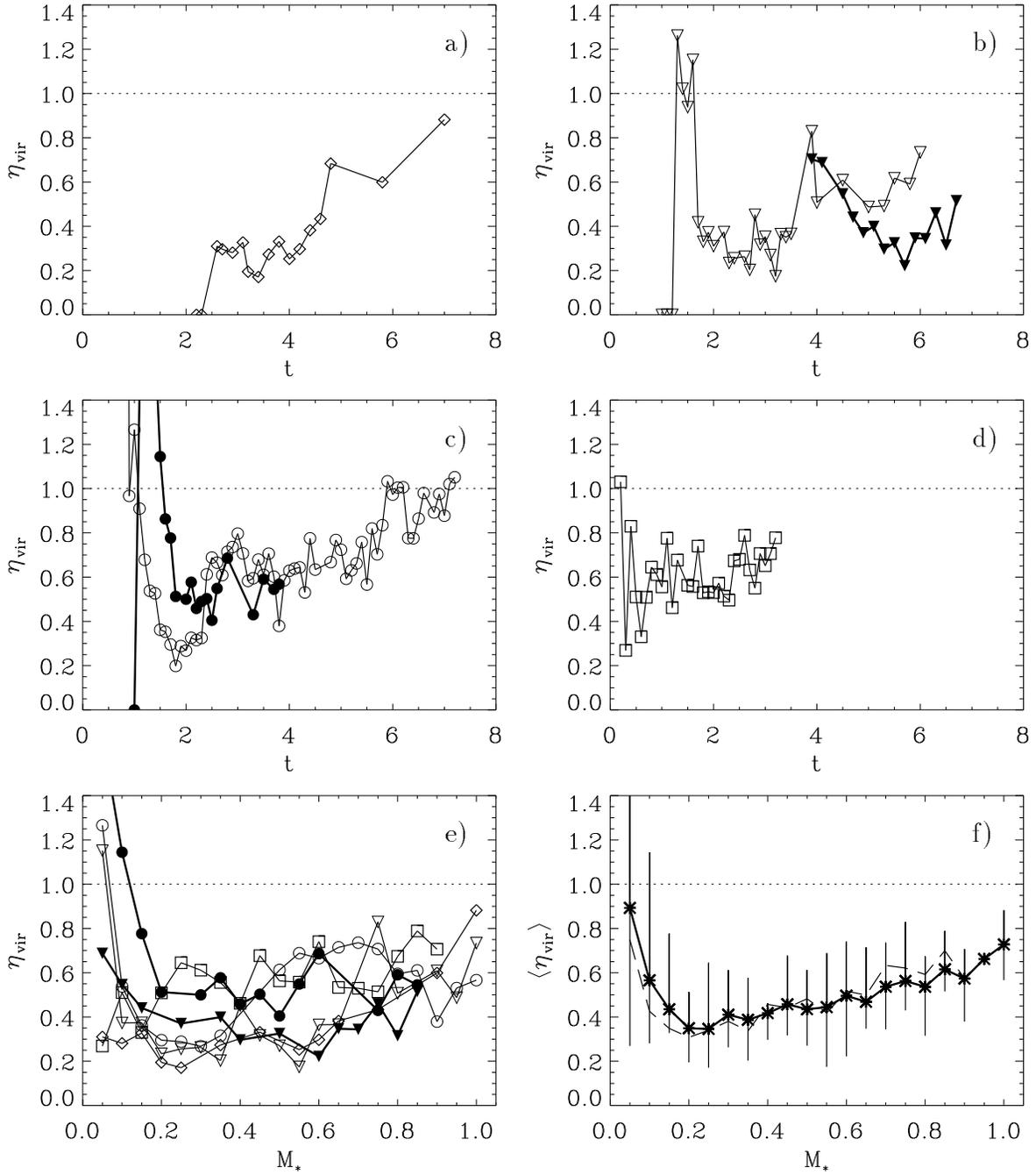}}
\end{picture}
\caption{\label{fig:energy-2} \footnotesize Time evolution of the virial coefficient
$\eta_{\rm vir}$: (a) for model ${\cal A}0a$, (b) for models ${\cal
A}1a$ and ${\cal A}1b$, (c) for models ${\cal A}2a$ and ${\cal A}2b$, and (d) for
model ${\cal A}3a$. The correspondence between each plotting symbol and
model is analogous to Fig.~\ref{fig:energy-1}. Plot of $\eta_{\rm
vir}$ as function of the total mass fraction $M_*$ accreted onto
protostellar cores (e) for each model individually and (f) averaged
over all models (thick solid line with error bars indicating the
statistical deviations) and averaged over the four models with
$2\times10^5$ particles (dashed line). }
\end{figure}

Figure \ref{fig:energy-1} specifies the energy and kinematic
properties of the protostellar cluster that forms in each of the above
models. In (a) the time evolution of the kinetic energy of the random
motions of the protostellar cores is plotted, $E_{\rm int}=1/2\sum_i
m_i ({\vec{v}}_i - {\vec{v}}_{\rm cm})^2$, where $m_i$ and
${\vec{v}}_i$ are the masses and velocities of individual cores $i$,
and ${\vec{v}}_{\rm cm}=\sum_i m_i {\vec{v}}_i/M_*$ is the
center-of-mass velocity of the protostellar cluster.  In (b) the
potential energy of the cluster disregarding the contribution from the
surrounding gas is shown, $E_{\rm pot}=\sum_{ij}Gm_im_j/r_{ij}$, with
the gravitational constant set to unity. The distance between two
cores $i$ and $j$ is denoted $r_{ij}=|{\vec{r}}_i - {\vec{r}}_j|$,
where each pair is counted once. The velocity dispersion $\sigma^2 =
\sum_i({\vec{v}}_i - {\vec{v}}_{\rm cm})^2$ and the total mass
accumulated in protostellar cores $M_*=\sum_i m_i$ are specified in
(c) and (d), respectively. In Fig.~\ref{fig:energy-1} and also in
Fig.~\ref{fig:energy-2}, open diamonds denote model ${\cal A}0a$ with
$\nu=0$, and triangles the models with $\nu=1$, ${\cal A}1a$ (open
symbols) and the high-resolution model ${\cal A}1b$ (filled symbols),
whose initial fluctuation spectrum is truncated at large scales,
i.e.~contains only modes with $4 \le k$. The two models with $\nu=2$,
${\cal A}2a$ and ${\cal A}2b$ are plotted with open and filled
circles, respectively. Finally, the model with the steepest initial
fluctuation spectrum (with $\nu=3$) is given by open squares. From
Fig.~\ref{fig:energy-1}, one immediately reads off the different
time delays for models with the different spectral slopes
$\nu$. Furthermore, one sees that the evolution of the energetic and
kinematic properties of the cluster in model ${\cal A}2a$ and ${\cal
A}2b$ is almost identical, independent of their different particle
number (as discussed in detail in Paper I). The delay between the two
models with $\nu=1$, ${\cal A}1a$ and ${\cal A}1b$, is due to the
truncation of the initial fluctuation spectrum of ${\cal A}1b$ as
discussed in \S~\ref{subsec:dependence-k}.

Figure \ref{fig:energy-2}, specifies the time evolution of the virial
coefficient, $\eta_{\rm vir} = 2 E_{\rm int}/|E_{\rm pot}|$. This
quantity is obtained from core properties only. The contribution of
the remaining gas in the cluster region is not included since we want
to estimate the possible dynamical evolution of the cluster {\em
after} a possible gas removal.  A value $\eta_{\rm vir} = 1$ suggests
that the cluster is in virial equilibrium, $\eta_{\rm vir} < 1$
indicates that gravitational attraction outweighs kinetic energy and
the cluster is contracting, and in the case $\eta_{\rm vir} > 1$ the
cluster will be expanding if the gas gets removed. To compare the
virial coefficient of each model at comparable evolutionary stages,
Fig.\ \ref{fig:energy-2}e plots $\eta_{\rm vir}$ as function of the
total fraction of gas converted into protostellar cores
$M_*$. Finally, Fig.~\ref{fig:energy-2}f specifies the average value
and its uncertainty.  The scatter at early times or equivalently at
very low values of $M_*$ is due to the small number statistics. As
more and more protostellar cores form, they follow the global gas flow
towards the common center of gravity, where they build up an embedded
dense cluster.  For simulations with an initial
$\nu\!=\!2$-fluctuation spectrum this has been discussed in detail in
Paper I. Our simulations show that the nascending protostellar
clusters are bound even without the gravitational contribution of the
gas. The conversion of gas into dense cores proceeds such that the
overall gain of potential energy is more or less balanced by the
increase of kinetic energy. However, due to the limited available gas
reservoir, the virial coefficient increases slowly with time in this
conversion process. This is best seen in Fig.~\ref{fig:energy-2}f. At
the time when the gas reservoir is completely depleted the virial
coefficient is $\eta_{\rm vir} \simeq 0.6$. Now the clusters behave
like collision-dominated $N$-body systems and develop a characteristic
core/halo structure with the virial coefficient approaching unity. The
overall velocity dispersion $\sigma$ and the energies $E_{\rm int}$
and $|E_{\rm pot}|$ decrease again (as visible in the evolution of
model ${\cal A}2a$ in Fig.~\ref{fig:energy-1}, open circles).

The situation becomes more complicated, when the subsequent collapse
of cores into into individual stars is considered. Whereas the cluster
of protostellar cores is a bound entity, this may no longer be true
for the resulting {\em stellar} cluster. If the star-formation
efficiency of individual cores is very low, a large fraction of the
gas accumulated in the protostellar core is expelled again (i.e.\ in
protostellar outflows) and will not end up in the stars. If this
mass-loss occurs rapidly, the resulting star cluster retains the
original velocity dispersion, while its potential becomes shallower
due to this mass loss, and as a consequence, the cluster
dissolves. Only for small or gradual mass loss can the system expand
adiabatically and remain bound (Geyer \& Burkert 2000).  Observations
(Motte, Andr\'{e} and Neri 1998, P.\ Andr\'{e} private communication)
and detailed numerical collapse simulations (Wuchterl \& Tscharnuter
2000, Wuchterl \& Klessen 2000), however, indicate that most of the
matter that gets accumulated in protostellar cores actually continues
to accrete onto the central protostellar object. Hence, our current
set of simulations is indeed indicative about the dynamical fate of
the resulting stellar cluster.

\section{Dependence on the Temperature}
\label{sec:dep-temperature}
For a given density, the Jeans mass is a function of the temperature
of the isothermal gas (Eqn.~\ref{eqn:def-jeans-mass}).  Changing the
temperature modifies the number of Jeans masses contained in
simulation. Since the isothermal models can be scaled to a wide range
of physical regimes, increasing the dimensionless temperature, i.e.\
the scaling parameter $\alpha$, but keeping the physical gas
temperature constant (say at 10K) is equivalent to zooming in more
closely onto smaller physical scales. Vice versa, decreasing the
scaling parameter implies to study a larger volume of molecular cloud
material.  For example, the behavior of systems with dimensionless
temperatures $\alpha=0.04$ is comparable to the dynamical evolution
within sub-regions of volume $1/8$ in a model with $\alpha=0.01$
(i.e.\ in one octant of the simulation cube).  This, however, neglects
the influence of large-scale modes that are present in the simulations
of larger volumes, and the approximation is only good for the first
stages of the evolution.  All simulations show that after one or two
global free-fall times the overall matter distribution is determined
by the growth of the largest possible mode in the system. There is
always {\em one} global minimum of the potential, towards which the
bulk of matter flows, leading to the formation of {\em one}
protostellar cluster.  The (periodic) boundary conditions of a model
with high temperature are not identical with the boundary conditions
of a subregion within a model with lower temperature, which contains a
comparable number of Jeans masses. As the time progresses in both
systems, the influence of the boundary conditions on the dynamical
evolution increases. At late times, deviations between both models are
expected and detailed comparison becomes less meaningful.  Properties
of the system that are strongly influenced by large-scale flows and
are the result of considerable dynamical evolution are expected to be
different, e.g.\ the spatial properties of the forming protostellar
clusters.  Properties that are less sensitive to these processes,
e.g.\ the overall slope of the clump-mass spectrum, are comparable in
both models at all times. This is true for the evolution of the mass
spectrum which is dominated by local processes, by the interplay
between gravitational attraction and gas pressure.

\begin{figure}[t]
\unitlength1.0cm
\hspace{-0.3cm}
\begin{picture}(16,11)
\put(-2.0,-9.0){\epsfbox{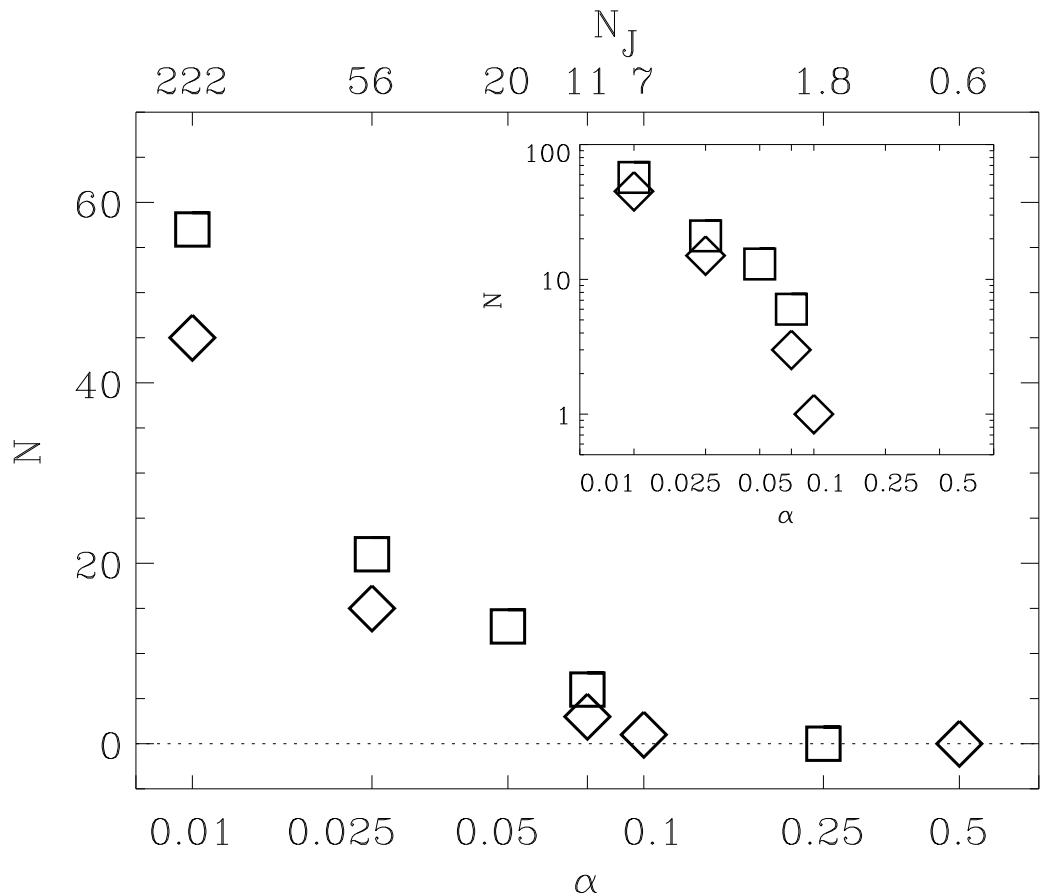}}
\end{picture}
\caption{\label{fig:number-of-sinks}\footnotesize  Total number $N_*$ of
protostellar cores that form during the dynamical evolution of the
isothermal models listed in Tab.~\ref{tab:models-T-variation} as
function of the temperature parameter $\alpha$. Open diamonds denote
the set of models with $\nu=1$ and open squares denote models with
$\nu=2$; dimensionless temperatures are given on a logarithmic
scale. The upper axis indicates the corresponding number of Jeans
masses $N_{\rm J}$ contained in the considered volume. The inlay gives
$N_*$ scaled logarithmically to indicate that the number of
protostellar cores declines with increasing temperature $\alpha$
roughly proportional to $\alpha^{-3/2}$. }
\end{figure}

To quantify the temperature dependence (i.e.\ the dependence on the
ratio of gravity forces to gas pressure forces), we perform a series
of simulations with varying temperature parameter $\alpha$, as listed
in Tab.~\ref{tab:models-T-variation}. All models contain $5\times10^4$
particles with different slopes of the initial fluctuation
spectrum. We consider $\nu=1$ and $\nu=2$. The covered temperature
parameters range from $\alpha=0.01$, corresponding to 222 Jeans
masses, to $\alpha=0.5$, which means the Jeans mass is larger than the
mass contained in the cube.  Figure~\ref{fig:number-of-sinks}
specifies the total number $N_*$ of protostellar cores that form
during the dynamical evolution as function of the temperature $\alpha$
(lower abscissa) or equivalently the number $N_{\rm J}$ of Jeans
masses in the system (upper abscissa). To indicate the power-law
behavior of the dependence of $N$ on $\alpha$, the inlay shows
$N(\alpha)$ in a log-log plot.  In the models with temperatures
$\alpha\sig0.25$ no fluctuations are massive enough in order to exceed
the Jeans limit. All initial density fluctuations are smeared out by
pressure forces and the system reaches a state of maximum homogeneity.
Collapse occurs only for models with lower temperatures or,
equivalently, regions that contain a sufficient number of Jeans
masses.  The number of forming protostellar cores scales roughly {\em
linearly} with the number of Jeans masses in the system $N \propto
N_{\rm J}$ or equivalently $N \propto \alpha^{-3/2}$. With our
definition of the Jeans mass (Eqn.~\ref{eqn:def-jeans-mass}) one core
typically forms for every four Jeans masses.

We find as a general feature of all our simulations of regions within
molecular clouds which have lost their turbulent support that their
subsequent dynamical evolution leads to the formation a cluster of
protostellar cores with a wide mass distribution. The protostellar
mass spectrum typically spans two orders of magnitude and always peaks
around the {\em average Jeans mass} of the system.  All of our
simulation cloud regions exhibit this behavior as long as they are
sufficiently massive to be able to form a cluster of stars. This is
{\em independent} of the adopted initial fluctuation spectrum. This
may seem somewhat surprising, as individual protostellar cores form
and grow in mass via a sequence of highly {\em stochastic} events
(Paper I). They form and feed from gas clumps and fluctuations of
widely different masses, sizes and density contrasts, each leading to
different collapse times and length scales. However, in a statistical
sense the overall dynamical evolution progresses such that the system
retains knowledge of its average properties (density and
temperature). Our numerical simulations therefore support the
hypothesis (e.g.\ Larson 1978, 1985) the {\em average Jeans mass} is
the main parameter that determines the peak of the IMF.

\section{Summary}
\label{sec:parameter-study-summary}
We have performed hydrodynamic simulations of the dynamical evolution
and fragmentation of regions within molecular clouds which have lost
their turbulent support leading to gravitational contraction. As the
most general approach which allows for an adequate description of the
statistical properties we have assumed that the density field
imprinted by the fading turbulent velocity field is described by
random Gaussian fluctuations. We vary the slope $\nu$ of the power
spectrum of the fluctuation field $P(k) \propto k^{-\nu}$ and the
number of Jeans masses contained in the considered volume. Spectra
with steep slopes ($\nu = 3$) correspond to initial matter
distributions that are highly fragmented and exhibit large density
contrasts. These systems are dominated by large-scale modes. In the
case of flat spectra ($\nu=0$), all spatial modes contribute equally
to the density field. Consequently, these systems appear more or less
homogeneous and exhibit no strong density contrasts. In Paper I we
concentrated on a detailed analysis of the case $\nu = 2$. Here, we
discussed the complete set of simulations with slopes ranging from
$\nu = 0$ to $3$ and we considered molecular cloud volumes that
contain a wide range of different masses relative to the average Jeans
mass (from $1\,M_{\rm J}$ to $222\,M_{\rm J}$).

The dynamical evolution of the simulated molecular cloud regions is
rather insensitive to the slope of the initial density fluctuation
spectrum. On a relatively short time scale, i.e.\ within one or two
free-fall times, the system evolves into a complex network of
intersecting filaments and collapsing clumps leading to the formation
of a compact cluster of accreting and interacting embedded
protostellar cores. These clusters form as bound entities, with
central densities being roughly a few $10^2$ times larger than the
initial average gas density. In their later phase of evolution,
however, the clusters dissolve due to close encounters between the
protostars (as discussed in more detail in Paper I).

In {\em all} simulations, the mass spectrum of collapsed cores is very
broad, has approximately log-normal shape and peaks roughly at the
average Jeans mass. This is independent of the adopted initial
conditions. The number of protostellar cores that form is linearly
proportional to the number of thermal Jeans masses in the studied
region.  Despite the fact that individual cores evolve through a
sequence of highly stochastical events, in a statistical sense the
evolution retains knowledge of the mean properties of the system.
This supports the hypothesis that the average Jeans mass is the main
parameter determining the peak in the stellar spectrum.

The current investigation and the detailed analysis presented in Paper
I demonstrates good agreement between the current isothermal
fragmentation model and the properties of observed star forming
regions.  This is somewhat surprising given the fact that we have
neglected magnetic fields, energetic feedback processes or external
environmental effects.  It indicates that the interplay between
self-gravity on the one side and thermal and turbulent pressure on the
other side is the dominant process that regulates the formation of
stellar clusters.

\acknowledgements We thank Rainer Spurzem and Pavel Kroupa for many
discussion on stellar clusters, and the anonymous referee for detailed
comments.




\begin{table}[p]
{\caption{Models with parameter $\alpha = 0.01$ and different initial power spectra
$P(k) \propto k^{-\nu}$ with $\nu=0 \dots 3$.}
\label{tab:models-N=0,1,2,3}}
\vspace{-0.25cm}
\begin{center}
\begin{tabular}[t]{lccccc}
\hline
{Model}    & {Power-Law} & {Temperature} & {Number of}        & {Particle}
& {Number of} \\
{Name$^a$} & {Exponent}  & {Parameter}   & {Jeans Masses$^b$} & {Number}
& {Collapsed Cores} \\
\hline
{~${\cal A}0a$}  & $\nu=0$ & $\alpha=0.01$ & $N_{\rm J} = 222$ &
$N=2 \times 10^5$ & $N_*= 45$ \\[0.2cm]
{~${\cal A}1a$}  & $\nu=1$ & $\alpha=0.01$ & $N_{\rm J} = 222$ &
$N=2 \times 10^5$ & $N_* = 49$ \\
{~${\cal A}1b^c$} & $\nu=1$ & $\alpha=0.01$ & $N_{\rm J} = 222$ &  $N=5\times 10^5$ & $N_* = 51$ \\[0.2cm]
{~${\cal A}2a^d$}  & $\nu=2$ & $\alpha=0.01$ & $N_{\rm J} = 222$ &  $N=2\times 10^5$ &  $N_* = 59$ \\
{~${\cal A}2b^d$} & $\nu=2$ & $\alpha=0.01$ & $N_{\rm J} = 222$ &  $N=5\times 10^5$ &  $N_* = 56$ \\[0.2cm]
{~${\cal A}3a$}  & $\nu=3$ & $\alpha=0.01$ & $N_{\rm J} = 222$ &  $N=2\times 10^5$ &  $N_* = 130$ \\
\hline
\end{tabular}
\end{center}
{\footnotesize{$^a$ The initial stages of all models have been
generated with the Zel'dovich method from a random homogeneous field
using $\delta t=1.5$ (for details see Paper I). }}\\
{\footnotesize{$^b$ Number $N_{\rm J}$ of thermal Jeans masses derived
from the average density in the simulated cube.}}\\
{\footnotesize{$^c$ The initial density distribution contains no
large-scale modes, the fluctuation spectrum used to generate this
model is truncated for wave numbers $k<4$. }}\\
{\footnotesize{$^d$ These two models are equivalent to models $\cal H$
and $\cal I$ in Paper I. }}
\end{table}

\begin{table}[p]
{\caption{ Models with initial power spectra $P(k) \propto k^{-\nu}$,
where $\nu=0$ and $\nu=2$, but different temperature parameters
$\alpha$.}
\label{tab:models-T-variation}}
\vspace{-0.25cm}
\begin{center}
\begin{tabular}[t]{lccccc}
\hline
{Model}    & {Power-Law} & {Temperature} & {Number of}        & {Particle}
& {Number of} \\
{Name$^a$} & {Exponent}  & {Parameter}   & {Jeans Masses$^b$} & {Number}
& {Collapsed Cores} \\
\hline
$~{\cal B}1a$ & $\nu=1$ & $\alpha=0.010$ & $N_{\rm J} = 222$ & $N=5\times 10^4$ &$N_* = 45$ \\ 
$~{\cal B}1b$ & $\nu=1$ & $\alpha=0.025$ & $N_{\rm J} = 56$ & $N=5\times 10^4$ &$N_* = 15$ \\
$~{\cal B}1c$ & $\nu=1$ & $\alpha=0.075$ & $N_{\rm J} = 11$ & $N=5\times 10^4$ &$N_* = 3$ \\
$~{\cal B}1d$ & $\nu=1$ & $\alpha=0.100$ & $N_{\rm J} = 7$ & $N=5\times 10^4$ &$N_* = 1$ \\
$~{\cal B}1e$ & $\nu=1$ & $\alpha=0.500$ & $N_{\rm J} < 1$ & $N=5\times 10^4$ &$N_* = 0$ \\
\hline
$~{\cal C}1a^c$ & $\nu=2$ & $\alpha=0.010$ & $N_{\rm J} = 222$ & $N=5\times 10^4$ &$N_* = 57$ \\
$~{\cal C}1b$ & $\nu=2$ & $\alpha=0.025$ & $N_{\rm J} = 56$ & $N=5\times 10^4$ &$N_* = 21$ \\
$~{\cal C}1c$ & $\nu=2$ & $\alpha=0.050$ & $N_{\rm J} = 20$ & $N=5\times 10^4$ &$N_* = 13$ \\
$~{\cal C}1d$ & $\nu=2$ & $\alpha=0.250$ & $N_{\rm J} = 2$ & $N=5\times 10^4$ &$N_* = 6$ \\
$~{\cal C}1e$ & $\nu=2$ & $\alpha=0.500$ & $N_{\rm J} < 1$ & $N=5\times 10^4$ &$N_* = 0$ \\
\hline
\end{tabular}
\end{center}
{\footnotesize{$^a$ The initial stages of all models have been
generated with the Zel'dovich method from a random homogeneous
field. For models ${\cal B}1d$ and ${\cal B}1e$ a shift interval $\delta
t=1.0$ is used. For ${\cal C}2a$ it is $\delta t = 2.0$ and for all
other ones $\delta t = 1.5$ (for details see Paper I). }}\\  
{\footnotesize{$^b$ Number $N_{\rm J}$ of thermal Jeans masses derived
from the average density in the simulated cube.}}\\
{\footnotesize{$^c$ This model corresponds to model $\cal A$ in Paper I.}}
\end{table}

\mbox{~~}
\newpage


\begin{figure}[p]
\unitlength1.0cm
\hspace{0.0cm}
\begin{picture}(16,18.5)
\put( 0.6, 0.0){\epsfbox{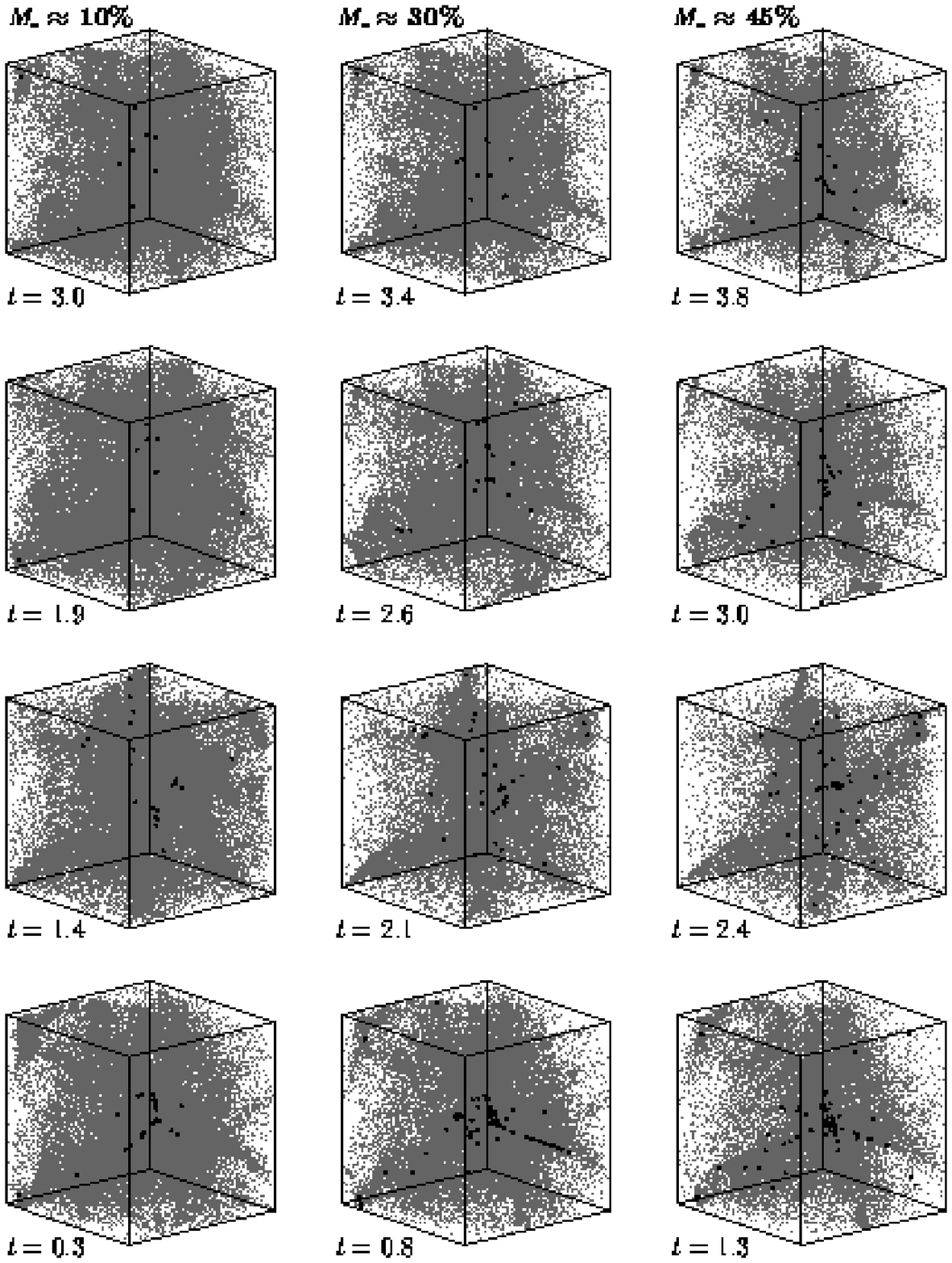}}
\put( 0.0,-0.7){\parbox[t]{16cm}{{Fig.~\ref{fig:cube-3D-T01Nx} ---
continued:} 
\footnotesize Comparison of the time evolution of
four models with different initial power spectra $P(k) \propto
1/k^\nu$: the distribution is plotted when $M_*\approx 10\,$\% (first
column), when $M_*\approx 30\,$\% (center column) and when $M_*\approx 
45\,$\% of the available gas is accreted onto protostellar cores.}}
\end{picture}
\end{figure}

\begin{figure}[tp]
\unitlength1.0cm
\hspace{0.0cm}
\begin{picture}(16,18.5)
\put( 0.6, 0.0){\epsfbox{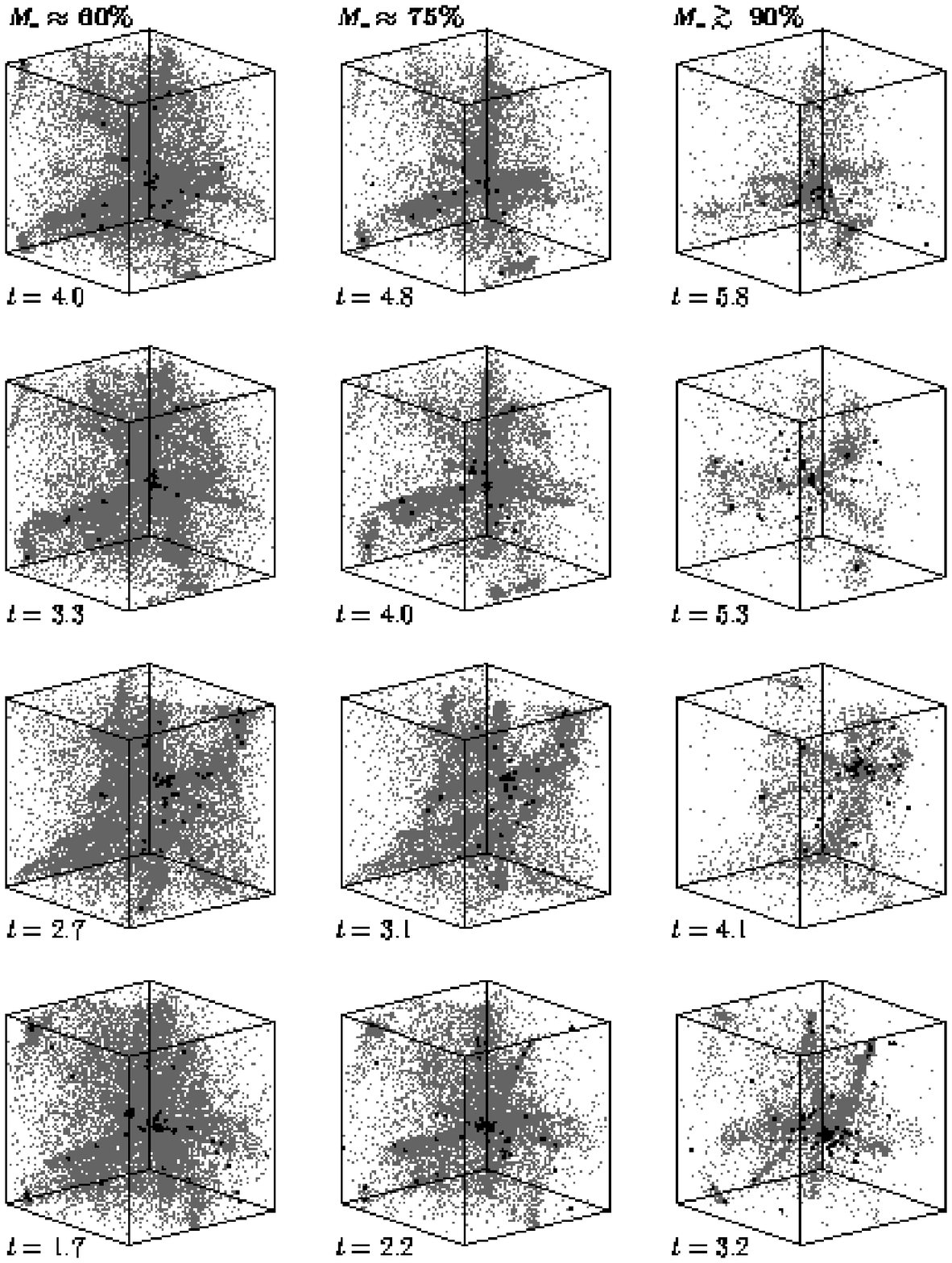}}
\put( 0.0,-0.7){\parbox[t]{16cm}{{Fig.~\ref{fig:cube-3D-T01Nx} --- continued:} \footnotesize Comparison of the time evolution of
four models with different initial power spectra $P(k) \propto
1/k^N$: the distribution is plotted when $M_*\approx 60\,$\% (first
column), when $M_*\approx 75\,$\% (center column) and when $M_*\sig
90\,$\% of the available gas is accreted onto protostellar cores.}}
\end{picture}
\end{figure}


\begin{figure}[p]
\vspace{-3.3cm}
\unitlength1.0cm
\begin{picture}(16,17.5)
\put(-2.5,-8.0){\epsfbox{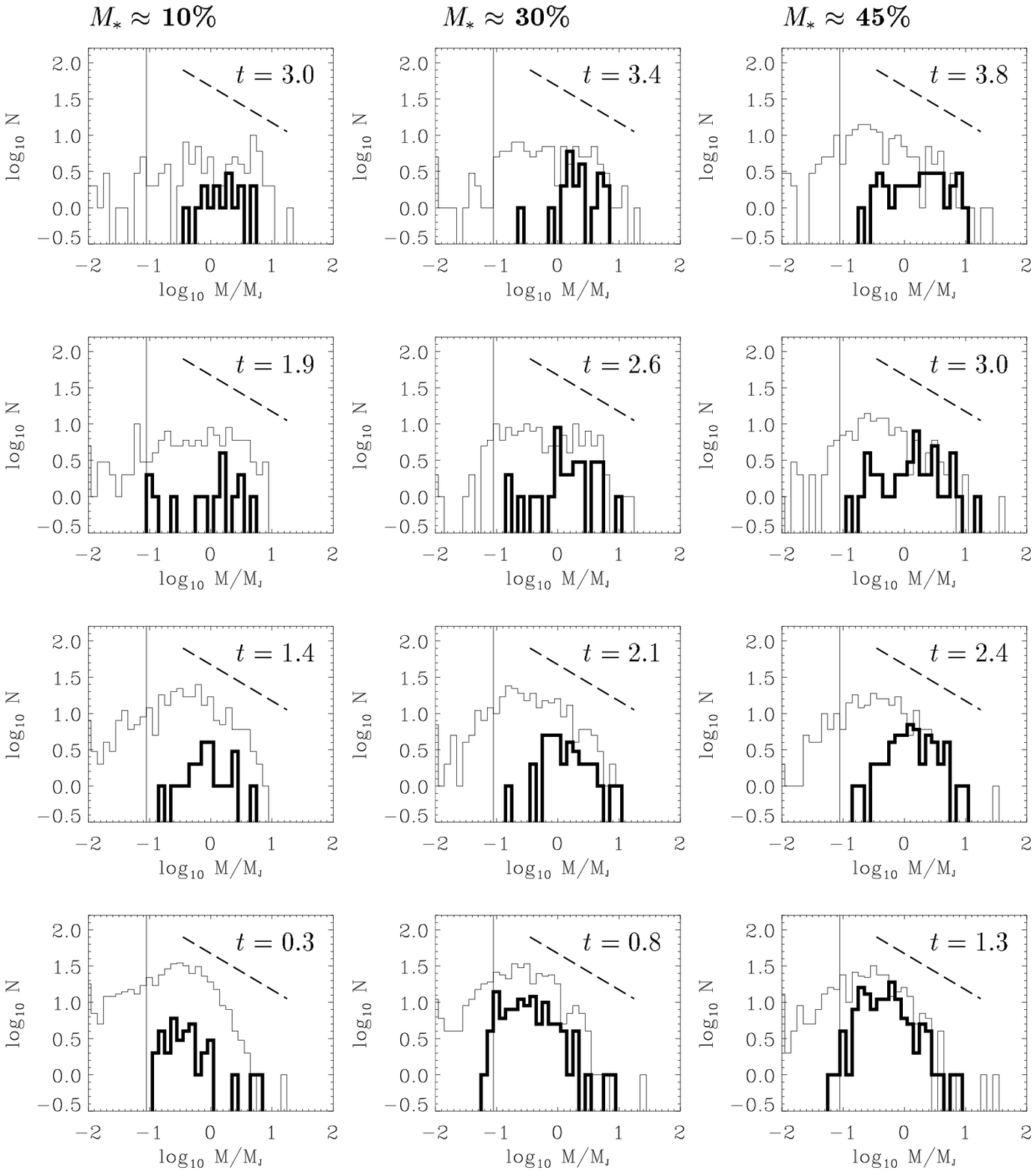}}
\put( 0.0,-1.5){\parbox[t]{16cm}{{Fig.~\ref{fig:mass-spectra-T01Nx} --- continued:} \footnotesize Mass
spectrum of gaseous clumps (thin lines) and protostellar cores (thick)
lines in models ${\cal A}0a$ (first row), ${\cal A}1a$ (second row), ${\cal A}2b$
(third row), and ${\cal A}3a$ (fourth row) at phases of the evolution when
$M_*\approx 10$\%, $M_*\approx 30$\%, and $M_*\approx 45$\%. }}
\end{picture}
\end{figure}

\begin{figure}[p]
\vspace{-3.3cm}
\unitlength1.0cm
\begin{picture}(16,17.5)
\put(-2.5,-8.0){\epsfbox{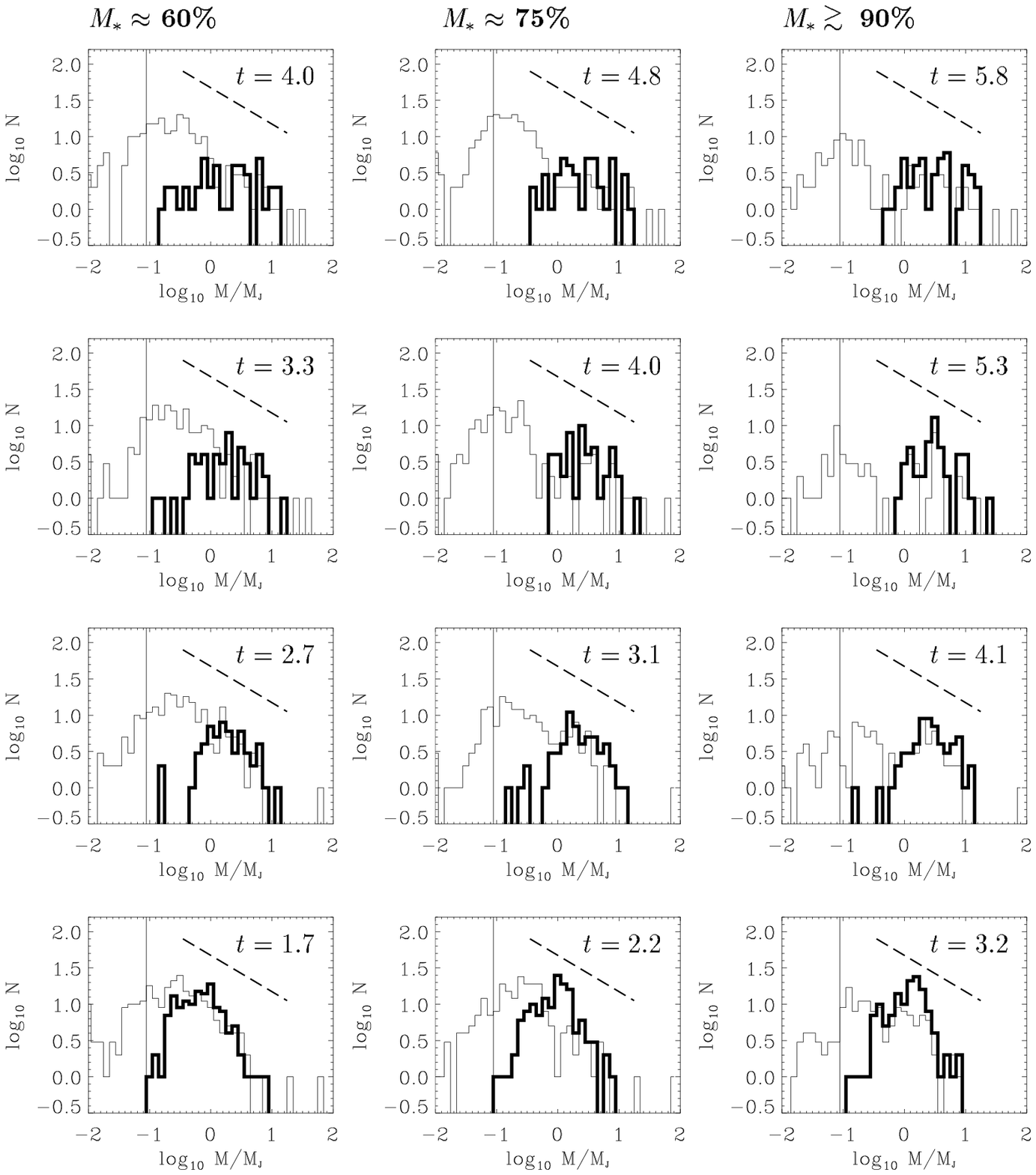}}
\put(0.0,-1.5){\parbox[t]{16cm}{ {Fig.~\ref{fig:mass-spectra-T01Nx}
--- continued:} \footnotesize Mass spectrum of
gaseous clumps (thin lines) and protostellar cores (thick) lines in
models ${\cal A}0a$ (first row), ${\cal A}1a$ (second row), ${\cal A}2b$ (third row),
and ${\cal A}3a$ (fourth row) at evolutionary phases when $M_*\approx
60$\%, $M_*\approx 75$\%, and $M_*\sig 90$\%.
}}
\end{picture}
\end{figure}

\end{document}